# Analysis of Beaulieu Pulse Shaping Family Based FIR Filter for WCDMA

*A.* S. Kang, Vishal Sharma

**Abstract-**The analysis and simulation of transmit and receive pulse shaping filter is an important aspect of digital wireless communication since it has a direct effect on error probabilities. Pulse shaping for wireless communication over time as well as frequency selective channels is the need of hour for 3G and 4G systems. The pulse shaping filter is a useful means to shape the signal spectrum and avoid interferences. Basically digital filters are used to modify the characteristics of signal in time and frequency domain and have been recognized as primary digital signal processing operations.

**Index Terms**— **FIR** filter, Beaulieu Pulse shaping, WCDMA.

——————————— ◆ ———————————

## 1. INTRODUCTION

With the recent exploding research interest in wireless communications, the application of signal processing to this area is becoming increasingly important. Indeed, it is the advances in signal processing technology that make most of today's wireless communications possible and hold the key to future services. Signal processing plays a crucial role in wireless communication for variety of applications [1-5]. Data communication using pulse shaping techniques has a critical role in a communication system. [6,7].

In digital telecommunication, pulse shaping is the process of changing the waveform of transmitted pulses. Its purpose is to make the transmitted signal suit better to the communication channel by limiting the effective bandwidth of the transmission. In Radio Frequency Communication, pulse shaping is essential for making the signal fit in its frequency band. [8-10]The application of signal processing techniques to wireless communications is an emerging area that has recently achieved dramatic improvement in results and holds the potential for even greater results in the future as an increasing number of researchers from the signal processing and communication areas participate in this expanding field. Due to intensive use of FIR filters in video and communication systems, high performance in speed, area and power consumption is demanded. Basically digital filters are used to modify the characteristics of signal in time and frequency domain and have been recognized as primary digital signal processing operations.

Digital Signal processing techniques are being used to improve the performance of 3G systems WCDMA (Wideband Code-Division Multiple Access), an ITU standard derived from Code-Division Multiple Access (CDMA), is officially known as IMT-2000 direct spread spectrum. W-CDMA is a third-generation (3G) mobile wireless technology that promises much higher data speeds to mobile and portable wireless devices than commonly offered in today's market. W-CDMA can support mobile/portable voice, images, data, and video communications at up to 2 Mbps (local area access) or 384 Kbps (wide area access). The input signals are digitized and transmitted in coded, spread-spectrum mode over a broad range of frequencies. A 5 MHz-

————————————————
- *A S Kang is Student ME (ECE)
  Deptt of Electronics and Communication Engg, University Institute of Engg and Technology, Panjab University,Chandigarh-160014.*
- *Vishal Sharma is lecturer in Deptt of Electronics and Communication Engg, University Institute of Engg and Technology, Panjab University,Chandigarh-160014.*

wide carrier is used, compared with 200 kHz-wide carrier for narrowband CDMA. The group delay plays a crucial role in pulse shaping digital finite impulse response filter. The value of group delay should be minimum for efficient performance of digital pulse shaping filter. Keeping in view the role of pulse shaping filter in wireless communication, the present study aims on analysis and simulation of pulse shaping digital finite impulse response filter for wideband code division multiple access to enhance its performance.

## 2. Analysis of Beaulieu Pulse shaping family:

Beaulieu pulse shaping family has been analyzed [11]. A flow chart has been prepared and a computer program has been written in MATLAB 7.3 version to compute the parameters for the analysis of Beauliu pulse shaping family. Flowchart for computation of parameters of Finite impulse response Pulse shaping Filter for WCDMA at 5 MHz (data rate 3.84Msymbols/sec) is shown in figure 1. The frequency responses for this family has been presented at half symmetry. The parameters of the flowchart are given as below:

$f_d$=Input Data rate

M=Interpolation factor/ Over sampling-Factor

$1/f_d$=Pulse Width

$f_s$= Sampling frequency

D=Group Delay

N=Filter Order

v=linspace(-500000,Fs,N);

y(:,k)=fef1(B,a1(k),v);

a1(k)=Roll off Factor;

i=linspace(-fs,fs,N);

y(:,k)=fsf(B,a1(k),i);

i=linspace(3840000,fs,N);

y( :k)=fisf(B,a1(k),i);

B=Bandwidth of 5MHz for WCDMA

### 3.1 Study of Effect of Roll Off Factor: 

To study the effect of variation of roll off factor from 0 to 1,the group delay is fixed at D=2 and Interpolation factor M=2 at input sampling data rate of 3.84Mbps(5MHz).Frequency Response

(Amplitude versus frequency(Hz), at roll off factor $\alpha$=0.1,0.5,1.0 is shown in figure 2

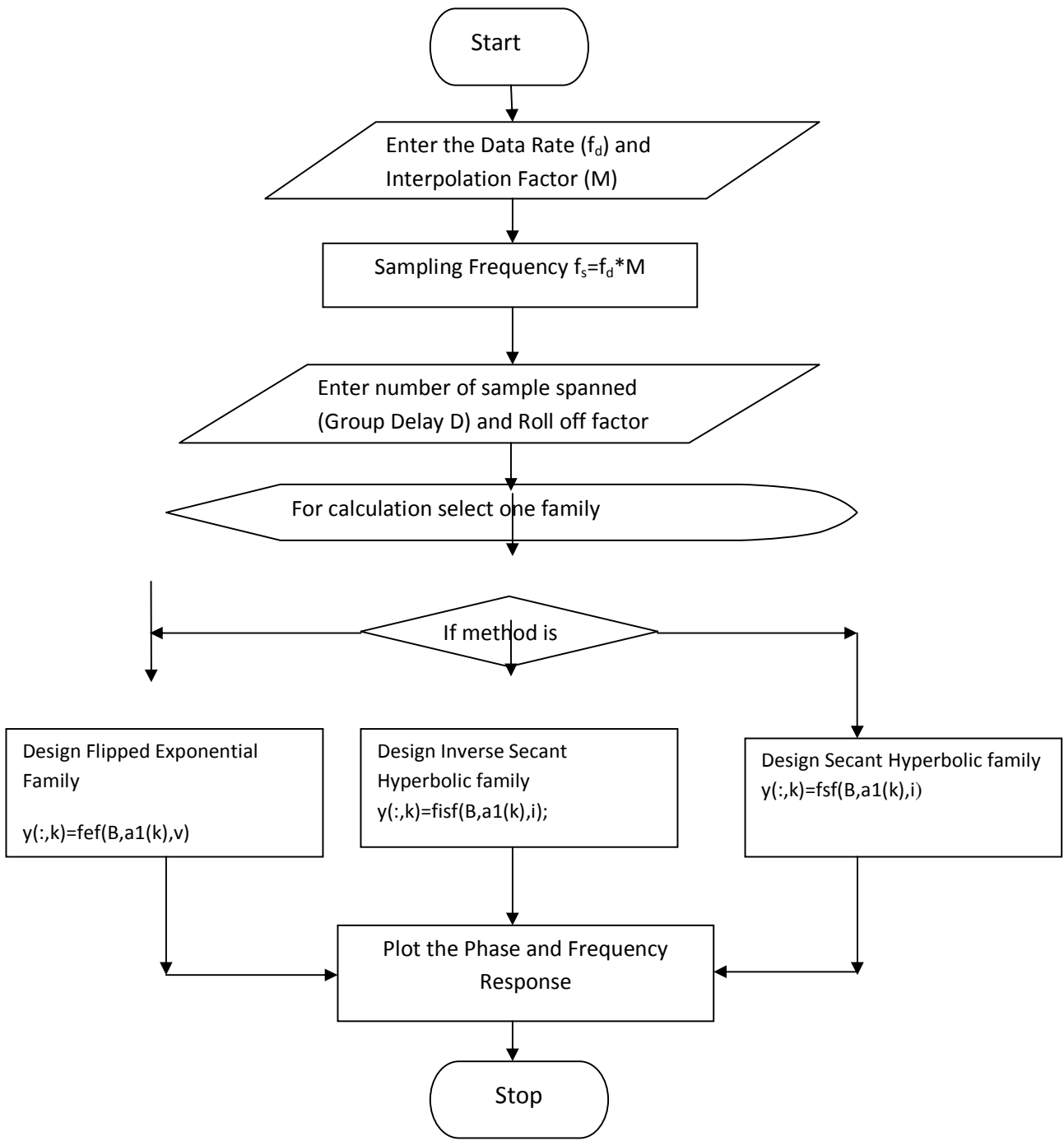

Figure 1. Flowchart for Pulse Shaping Families

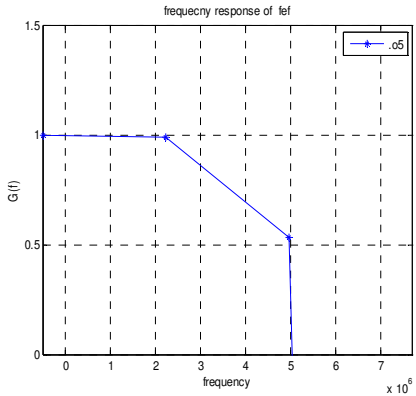

Alpha=0.1

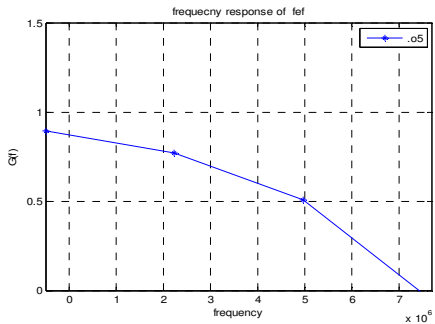

Alpha=0.5

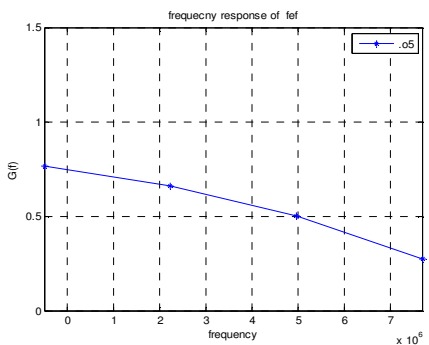

Alpha=1

Figure 2 Frequency Responses for fexp at Alpha 0.1 to 1.0

In the case of **frequency domain representation** of flipped exponential pulse shaping filter, ,amplitude changes from 1 to 0.8 as roll off factor alpha changes from 0.1 to 0.5.The amplitude further decreases from 0.8 to 0.75 as roll off factor is changed from 0.5 to 1. So, with increasing frequency range for flipped exponential pulse, the amplitude goes on decreasing. It has been observed that main lobe also expands from 5MHz to7.5MHz and then beyond 8MHz as the roll off is increased from 0.1 to 0.5 and then from 0.5 to 1 which shows side lobe tail attenuation occurs with delay. It is clear from figure 2. The passband is found to increase. The Responses[ Magnitude w.r.t Normalised Frequency and Phase w.r.t Normalised Frequency]for Flipped Exponential Pulse at fix value of Interpolation Factor M=2 and Group Delay D=2 at 5MHz BW with input signal sampling data rate of 3.84Msymbols/second at roll off factor=0.1,0.5,1.0 is shown in figure 3

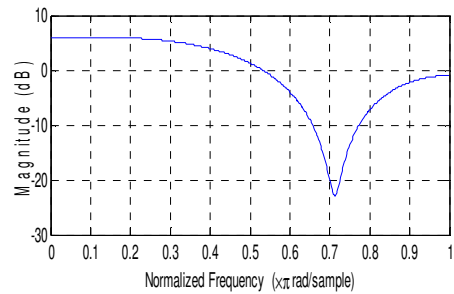

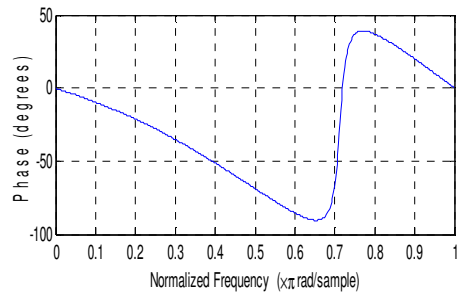

Alpha 0.1

towards alpha0.5 and then to alpha1.0, the stop band starts at normalized frequency of 0.67 and 0.55 The phase response w.r.t normalized frequency is found to vary 0 to -80degree at normalized frequency of 0.65 which then changes to -60 at normalized frequency of 0.55 on changing alpha from 0.1 to 0.5.It was found that phase response changed to -65 degrees at normalized frequency 0.4 on increasing alpha from 0.5 to 1.

### 3.2 Study of Effect of Variation of Group Delay D

To study the effect of variation of group delay D from 2 to 10,the roll off factor fixed at 0.22 and Interpolation factor M=2 at input sampling data rate of 3.84Mbps(5MHz).Different Frequency Response (Amplitude versus Frequency(Hz), at D=2,4,6,8,10 is shown in figure 4.

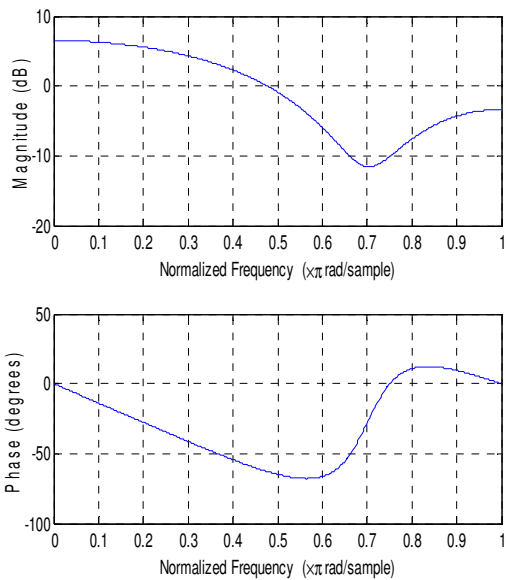

Alpha=0.5

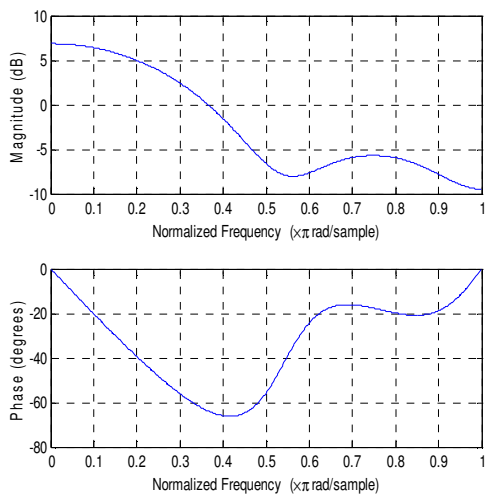

Alpha=1

Figure 3 Magnitude and phase response w.r.t to normalized frequency for fexp family at different Alpha

The magnitude is found to vary from 6 to-22 at normalize frequency of 0.72 and then stop band attenuation occurs at alpha 0.1.As we proceed

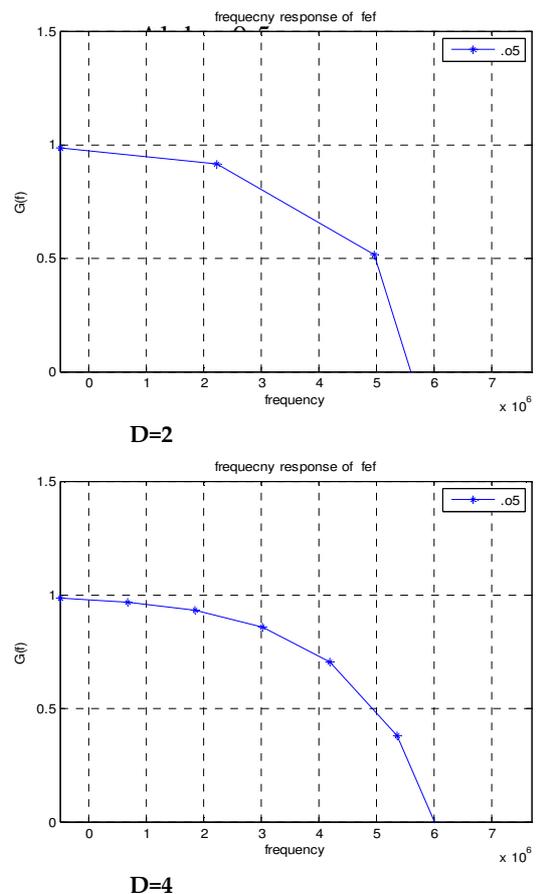

D=2

D=4

expands from 5.5 MHz to 6MHz. It means side lobe tails suppression occurs with delay.

The magnitude versus normalized frequency and phase versus normalized frequency responses for fexp pulse at different values of D are shown in figure 5 and 6 respectively.

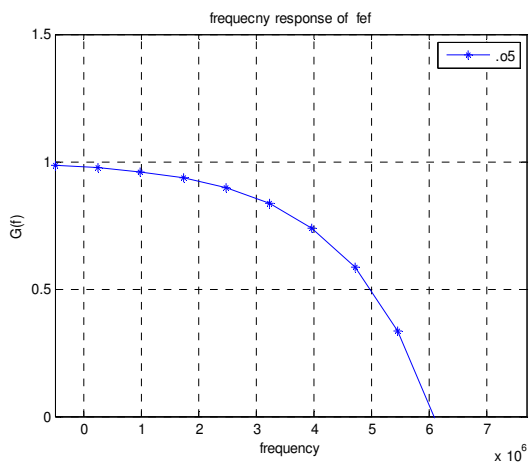

**D=6**

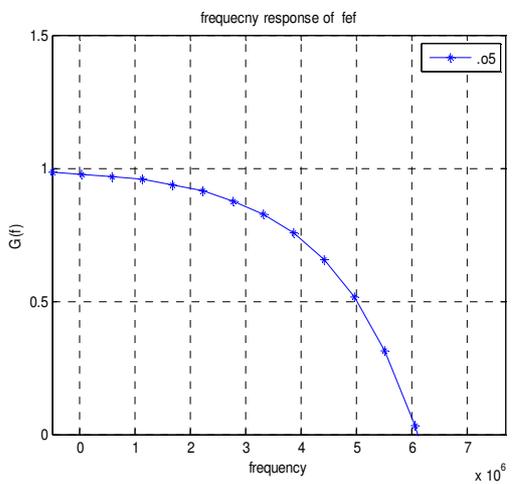

**D=8**

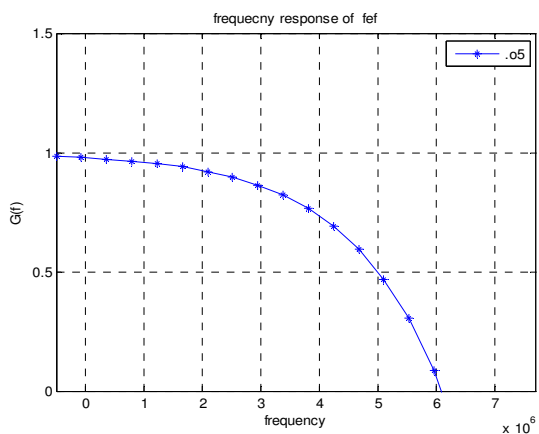

**D=10**

Figure 4. Frequency Responses for fef at different values of D

It is clear from figure 4 that with increasing D values from 2 to 10, the magnitude (amplitude) remains at 1 . On successively increasing D values, main lobe

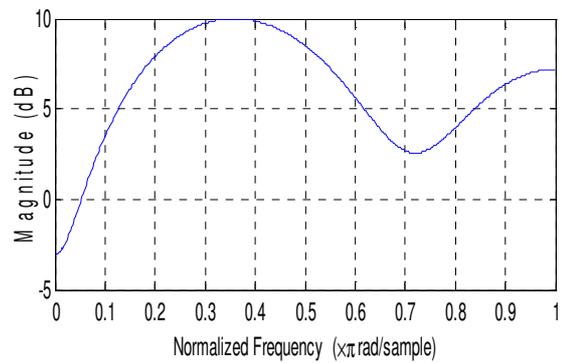

**D=2**

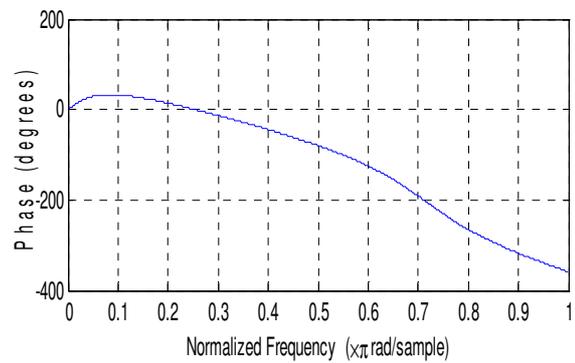

**D=4**

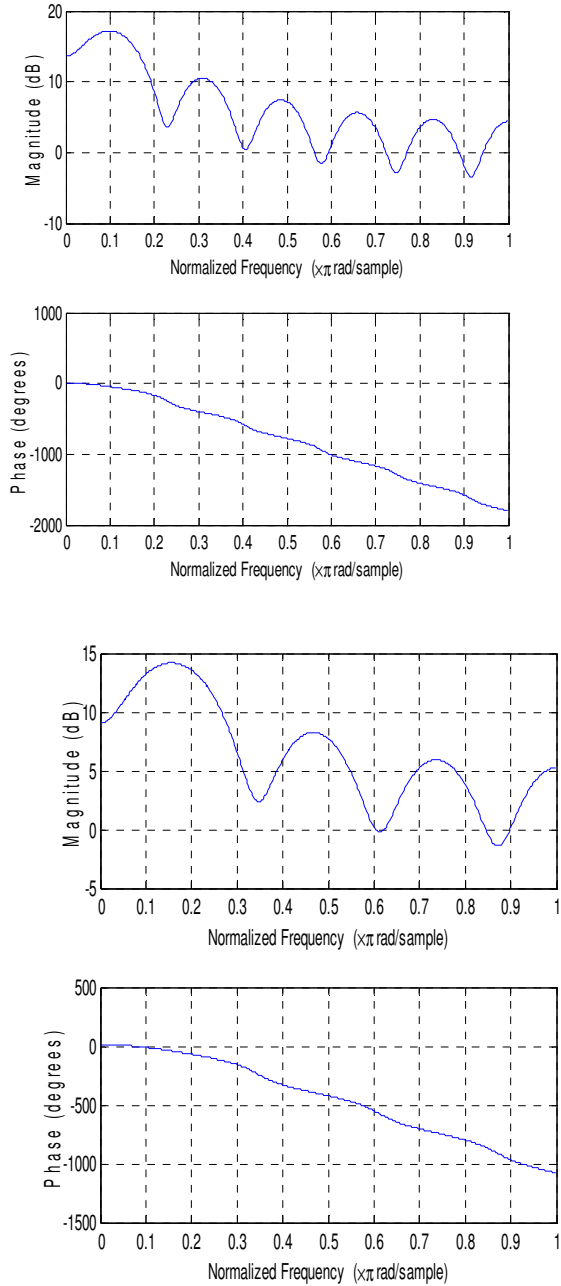

**D=6**

Figure 5 Magnitude and phase response w.r.t to normalized frequency for fexp family at Group Delay (D=2, 4 and 6)

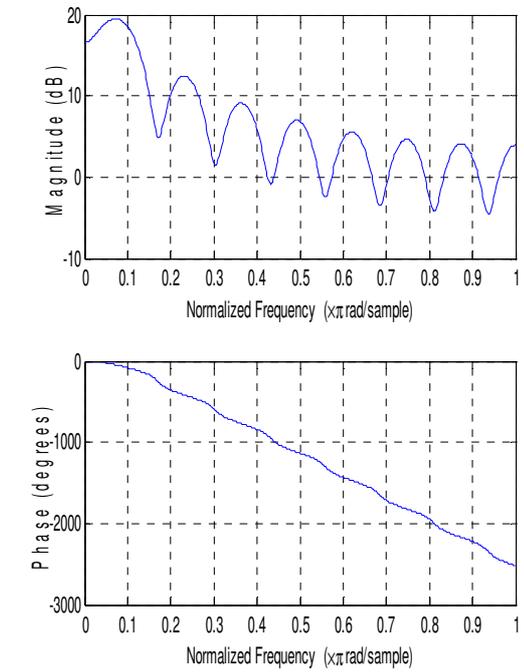

**D=8**

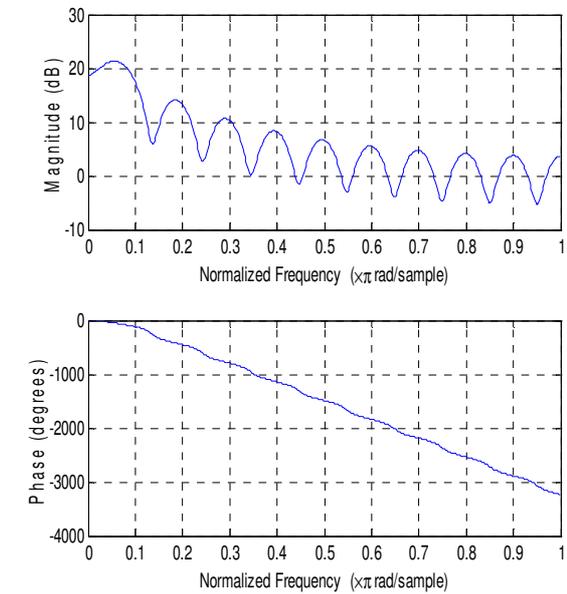

**D=10**

Figure 6 Magnitude and phase response w.r.t to normalized frequency for fexp family at Group Delay at 8 and 10

The magnitude response w.r.t normalized frequency is found to shift from -3 to 10 at normalized frequency of 0.34 at D=2.As D is increased from 2 to 4, the magnitude shifts from 9.8 to 3 at normalized frequency of 0.34.The magnitude shifts from 12 to 4 at normalized frequency of 0.22 in case of D=6.At D=8, magnitude decreases from 16 to 5 at normalized frequency of 0.16.At D=10, magnitude deceases from 20 to 7 at normalized frequency of 0.13.

At D=2,phase w.r.t normalized frequency decreases from 0 to -370 degrees at normalized frequency of 1.At D=4,phase w.r.t normalized frequency decreases from 0 to -1100 degrees at normalized frequency greater than 1.At D=6,phase w.r.t normalized frequency decreases from 0 to -1800degrees at normalized frequency of 1.At D=8,phase w.r.t normalized frequency decreases from 0 to -2500degrees at normalized frequency greater than 1.At D=10,phase w.r.t normalized frequency decreases from 0 to -3200degrees at normalized frequency greater than 1.So with increasing D, phase decreases with increasing values of normalized frequency. In magnitude versus normalize D frequency, stop band attenuation occurs early at decreasing values of normalized frequency as D is increased from 2 to 10 [12].

## 3.3 The study of effect of variation of Interpolation factor M

The study of effect of variation of Interpolation factor M keeping roll off factor 0.22 and group delay D fixed at 2 at 5MHz bandwidth has been studied at M(Even) Different Responses (Amplitude versus frequency(Hz))at M=2,4 and 6 are shown in figure 7.

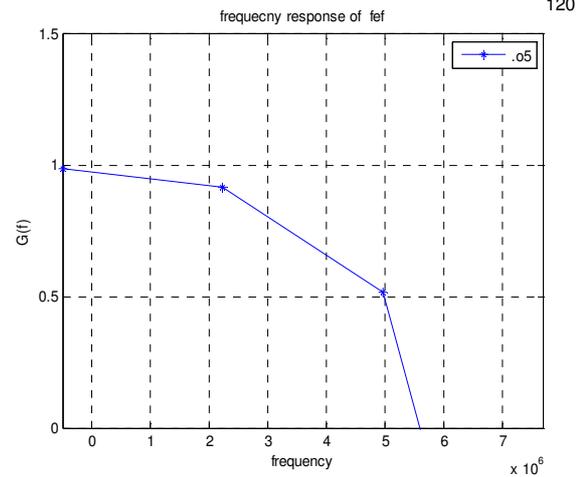
M=2

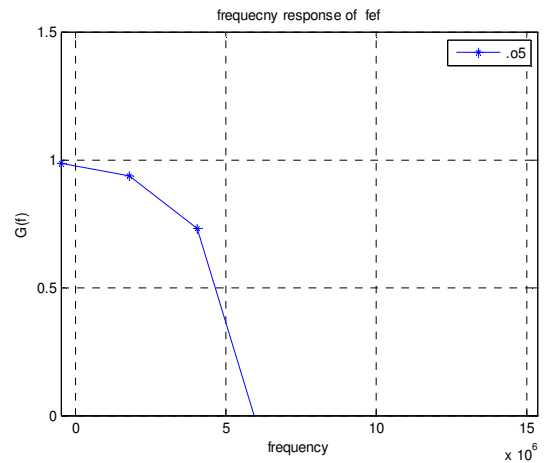
**M=4**

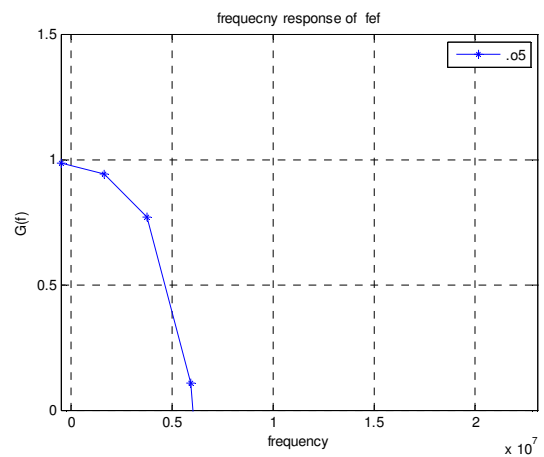
**M=6**

Figure 7 Frequency Responses for fef at different even M values

The value of M is varied from 2 to 4 and then from 4 to 6, it has been observed that with even values of M, amplitude remains same at 1 in amplitude versus frequency response representation.

The same study has been carried out odd Values of M for Flipped exponential pulse. The amplitude versus frequency responses at M=3, 5, 7 are shown in figure 8.

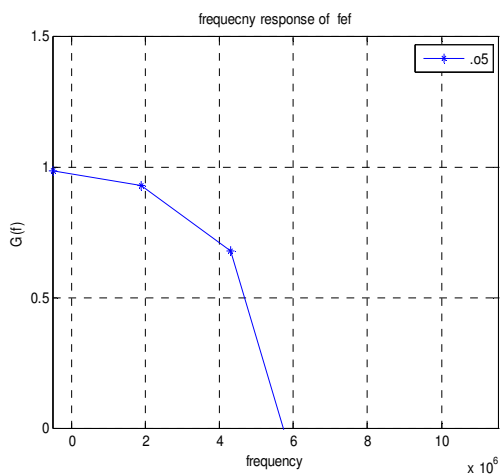

**M=3**

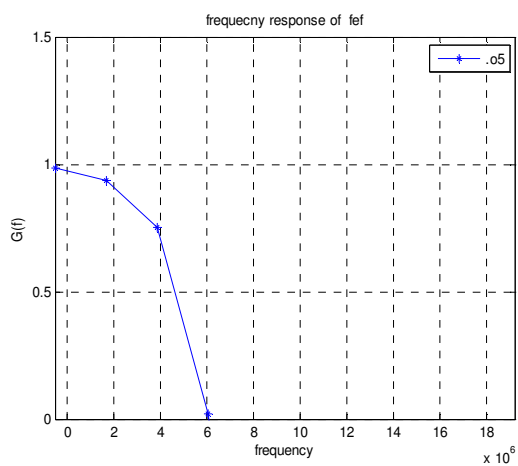

**M=5**

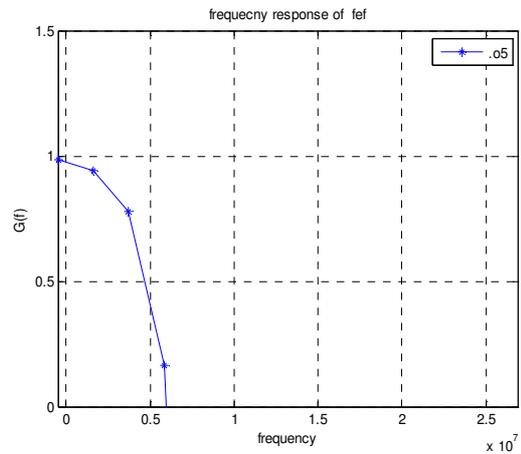

**M=7**

Figure 8. Frequency response of fexp family at different odd values of M

It is clear from figure 8 that in case of increasing the odd M Values, magnitude remains same at 1 with frequency changing from 5.8 to 6.i.e.it shows expansion trend in main lobe. Magnitude versus Normalised Frequency and phase versus Normalised frequency for M(Even=2,4,6) at D=2, alpha=0.22 are shown in figure 9.

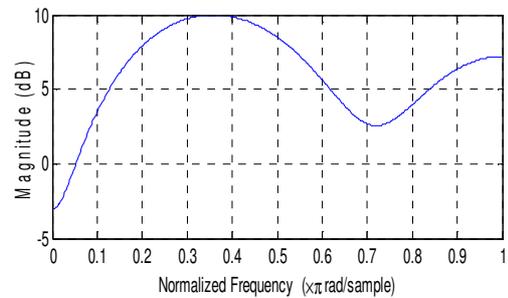

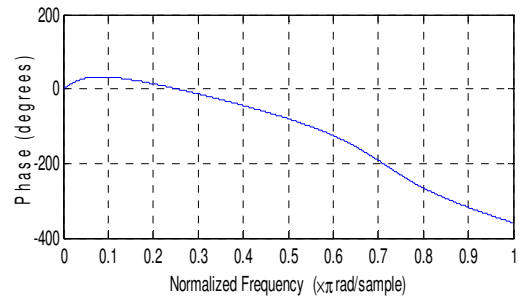

**M=2**

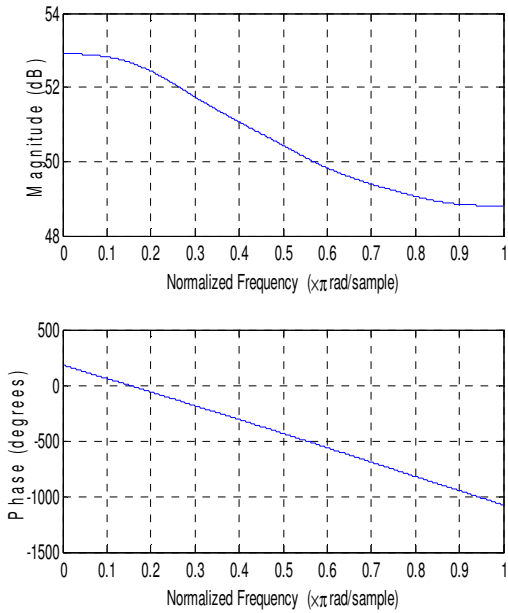

**M=4**

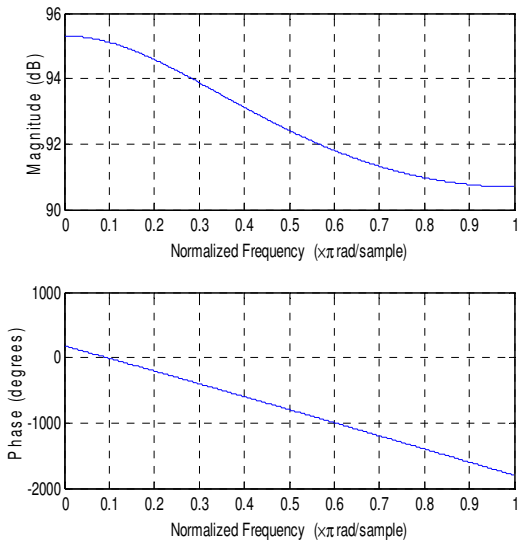

**M=6**

Figure 9. Magnitude and Phase response w.r.t normalized frequency of fexp at even M values

Magnitude with respect to normalized frequency shifts from -3 to 10 at normalized frequency of 0.32 at M=2.For M=4,magnitude w.r.t normalized frequency shifts from 53 to 49 at normalized frequency of 1.At M=6,magnitude w.r.t normalized frequency shifts from 95.2 to 90.7 at normalized frequency 1. Phase w.r.t Normalised frequency shifts from 0 to-390 at normalized frequency of 1 at M=2 For M=4,phase w.r.t normalized frequency shifts from 100 to -1050 at normalized frequency of 1.For M=6,phase w.r.t normalized frequency decreases from 100 to -1900 at normalized frequency of 1.So,it is clear from figure 9 phase is found to be decreased as the normalized frequency is increased. At lower values of M(Even)in the magnitude versus frequency plot,sidelobe occurs at early normalized frequency value.

The similar study has been carried for different odd values of interpolation factor the value of M is varied from 3 to 5 and then from 5 to 7. The different responses are given in figure 10.

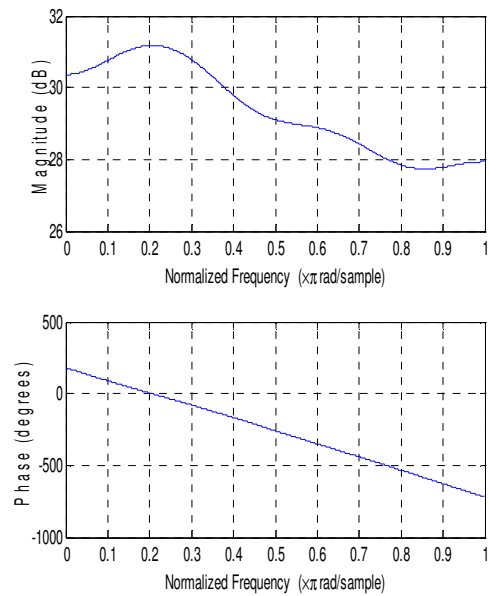

**M=3**

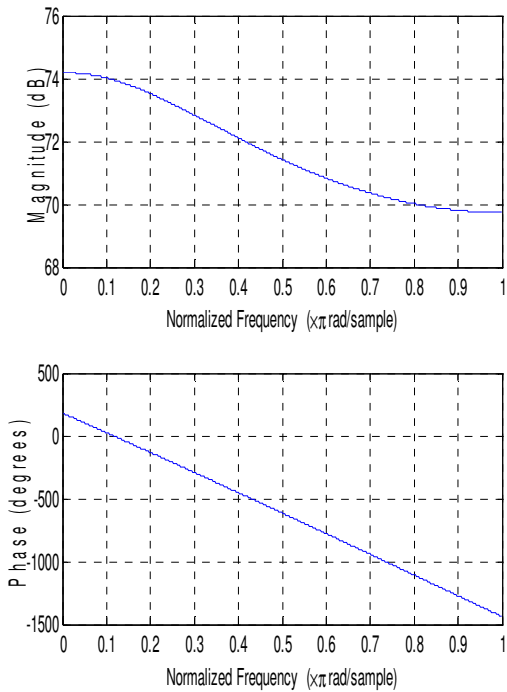

**M=5**

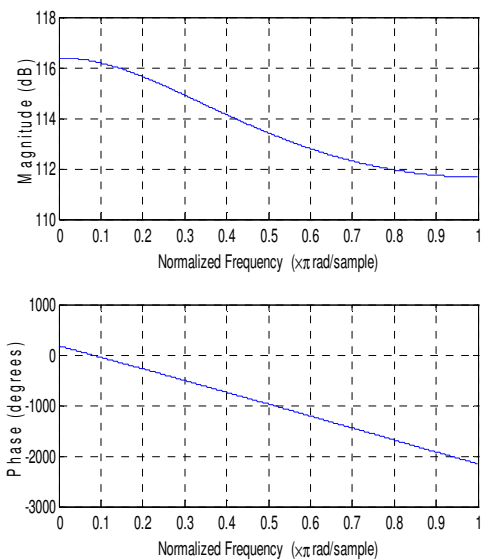

**M=7**

Figure 10. Magnitude and Phase response w.r.t normalized frequency of fexp at odd M values

Magnitude w.r.t Normalized frequency decreases from 30.5 to 28 at normalized frequency of 1 at M=3.At M=5, magnitude w.r.t normalized frequency decreases from 74.4 to 69.8 at normalized frequency of 1.At M=7, magnitude w.r.t normalized frequency decreases from 116.5 to 111.8 At M=3, phase angle w.r.t normalized frequency decreases from 100 to -700degrees at normalized frequency of 1. At M=5, phase w.r.t normalized frequency decreases from 100 to -1490degrees at a normalized frequency of 1.At M=7,phase w.r.t normalized frequency decreases from 100 to -2200 at normalized frequency of approximately 1.So ,Magnitude w.r.t normalized frequencies decreases with increase in M odd values. Also phase w.r.t normalized frequency decreases with increasing M odd values [13-15].

**Conclusion:** The group delay plays a crucial role in pulse shaping digital finite impulse response filter. The value of group delay should be minimum for efficient performance of digital pul;se shaping filter[12-20].The present study has highlighted the role of Beaulieu pulse shaping filter in WCDMA. The time and frequency response of square root raised cosine pulse shaping filter at 5MHz bandwidth has been studied .The effect of variation of Roll off factor alpha has been studied in case of amplitude versus frequency response, magnitude w.r.t normalized frequency and phase w.r.t normalized frequency. The effect of variation of group delay D i.e. number of symbols spanned by impulse response is studied at fix value of alpha=0.22 as well as at fix value of interpolation .The effect of variation of interpolation factor M has also been studied at roll off factor 0.22 and different values of D. At fix value of roll off factor0.22, there should be a tradeoff between D and M for better performance of pulse shaping filter for WCDMA based wireless communication system.

**Acknowledgement**: The first author is thankful to **Dr B S Sohi, Director UIET,Sec-25,PANJAB UNIVERSITY Chandigarh** for discussion and valuable suggestions during technical discussion at WOC conference at **PEC(Panjab Engg College) Chandigarh** during my presentation of research paper, "Digital Processing and Analysis of Pulse Shaping Filter for wireless communication "The help rendered by **S Harjitpal Singh,a research scholar** of **NIT Jalandhar** as well as **Mr Hardeep Singh,Research Scholar**, Communication Signal Processing research Lab, Deptt of Electronics Technology **GNDU Amritsar** is also acknowledged.

**AuthorBiography**

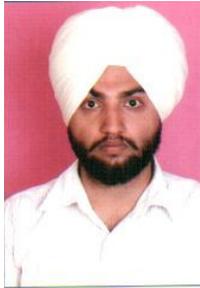

**Er A S Kang** received **B.Tech** degree in Electronics & Communication Engg from Deptt of ElectronicsTechnology,**GNDU Amritsar**,India.He received his **Masters** degree in Electronics and Communication Engg(with University Merit Certificate) from University Institute of Engg & Technology,**Panjab University,Chandigarh**,India.Presently ,he is working as Assistant Professor in Deptt of Electronics & Communication Engg at SSG Panjab University Regional Centre,Hoshiarpur.He has **two research publications in International Journals and eight publications in International** Conferences of repute among which the publications at **IEEE ISM2008at IISc Bangalore**(INDIA) and **IEEE IACC09** atn **Thapar University,Patiala**,and WECON2008 at **PEC Chd**,**IEEE SIBCON at RUSSIA** are commendable.His research interests include Wireless and Mobile Communication ,Signal processing in Wireless Communication.